\documentstyle[12pt]{article}
\begin{document}

\def\calB{{\cal B}}
\def\calM{{\cal M}}
\def\calR{{\cal R}}
\def\calS{{\cal S}}

\def\beq{\begin{equation}}
\def\eeq{\end{equation}}

\begin{titlepage}
\vspace{3cm}
\begin{flushright}
UT-891 \\

May 2000
\end{flushright}
\vskip 0.8cm
\begin{center}
{\Large \bf Action of Singular Instantons of Hawking--Turok Type}
\end{center}
\vskip 1cm
\begin{center}
Masashi Hashimoto
\vskip 1cm
{\em Department of Physics,\\
University of Tokyo, \\
Tokyo 113-0033, Japan}\\ 
(May 11, 2000)

\end{center}
\vskip 1cm

\thispagestyle{empty}

\begin{abstract}
Using Kaluza--Klein technique we show that the singularity of 
Hawking--Turok type has a fixed point (bolt) contribution 
to the action in addition to the usual boundary contribution. 
Interestingly by adding this contribution 
we can obtain a simple expression for the total action which is 
feasible for both regular and singular instantons.
Our result casts doubt on the constraint proposed by Turok in 
the recent calculation in which Vilenkin's instantons are 
regarded as a limit of certain constrained instantons.

\end{abstract}
\end{titlepage}

\section{Introduction}   
$\quad \, $ Hawking and Turok \cite{HaT98Feb, TuH98} found that a 
family of gravitational instantons exist for a generic potential 
for the scalar field. In the Hartle--Hawking `no-boundary' 
proposal \cite{HaH83} the universe is supposed to be created 
from a compact instanton, 
so that Hawking--Turok instantons 
have a significant importance. 
The novelty is the allowance of singularities of special kind 
(Hawking--Turok 
(HT)-singularity) 
in the curvature and the 
scalar field. Hawking and Turok argued that the use of such singular 
instantons is justified because of the finiteness of the action. 
In addition, the singularity is mild and behaves as a reflecting boundary 
for scalar and tensor cosmological perturbations \cite{HaT98Feb, 
Gar98Mar, GMST98, Gar99}.
Hence the Cauchy problem seems to be well posed and the model is 
well suited for the quantisation of small perturbations and for 
comparison with observations. As another justification, 
Wu \cite{Wu98} argued that Hawking--Turok instantons may be regarded as 
`constrained gravitational instantons,' whose action 
is required to be stationary under the constraint that the 3-geometry 
is given on the three-surface where quantum transition occurs. 

On the other hand, there is a criticism by Vilenkin \cite{Vil98} that 
asymptotically flat HT-singular instantons would destabilise the flat 
spacetime. Among such instantons especially the ones which do not 
have the potential for the scalar field are called as Vilenkin's 
instantons. To this criticism it was argued \cite{Gar98Apr} that 
Vilenkin's instanton can be taken as compactified five-dimensional 
Schwarzschild metric and the large extra dimension would 
metastabilises the flat spacetime. 
This mechanism was first discussed by Witten \cite{Wit82}.
Turok \cite{Tur99} argued that with a careful definition of a 
constraint Vilenkin's instantons possess no negative mode, so 
that they do not lead to the decay of flat spactime.

With such arguments 
it is desirable to examine the contribution of the HT-singularity 
to the action with much care. 
Most simply 
it is calculated as follows \cite{Vil98, BoL98}: at first we 
consider a manifold $\calM^{\prime}$
which consists of the whole spacetime $\calM$ except a four-sphere 
which contains the singular point and then make the excised 
four-sphere shrink. In the 
zero-volume limit 
of the excised four-sphere, the Gibbons--Hawking boundary term 
\cite{Yor, GiH} of the
boundary of $\calM^{\prime}$ would give the contribution of the 
singularity. 
Other than this simple method, there are several attempts 
which do not agree with
each other.
Garriga \cite{Gar98Mar} calculated the contribution of 
the singularity by regularizing the singularity with a membrane 
which wraps 
the singularity and found that the contribution of the singularity 
is just $1/3$ of the Gibbons--Hawking term. But the method was criticised 
by Bousso and Chamblin \cite{BoC} because of allowing negative energy density. 
In ref.\ \cite{BoC} it is considered to regularise the singularity with 
a membrane of a positive energy density which couples to four-form 
field strength. 
Garriga calculated in ref.\ \cite{Gar98Apr} the action of 
Vilenkin's instanton by 
showing that compactified five-dimensional Schwarzschild metric looks 
as Vilenkin's instanton and obtained the value of the $1/3$ of 
the Gibbons--Hawking term.
In ref.\ \cite{Gon98} Gonz\'{a}lez-D\'{\i}az calculated the 
action of Vilenkin's instanton which is replaced the singularity 
with an axionic wormhole and obtained the value zero.
Turok \cite{Tur99} considered to define Vilenkin's instantons 
as a limit of constrained instantons which are non-singular and have no 
boundary, and obtained the value of the whole Gibbons--Hawking term 
as the value of the action. 

Although the HT-singularity has been often regularised with new matter
in these calculations, 
the singularity at the transition surface from Euclidean to
Lorentzian metric would be allowed 
if we take HT-singular instantons as constrained gravitational
instantons. On the other hand  the simplest method mentioned 
above to calculate the
contribution of the HT-singularity would not be 
justified because the contribution of the excised spacetime is overlooked.  
For example, in the case of the Euclidean Schwarzschild metric 
the limiting value of Gibbons--Hawking term does not contribute to the 
action although it has a non-zero value. For that reason let us 
consider to calculate the contribution of the excised 
HT-singularity. To do this
we review the case of the Euclidean Schwarzschild metric in 
the next section. 
Then we apply it to the case of the HT-singular instantons 
using Kaluza--Klein technique.

\section{The case of Euclidean Schwarzschild metric: A simple example}

$\quad \, $ The Lorentzian Schwarzschild metric is given by 
\beq
	ds^2 = - \left( 1-\frac{2MG}{r} \right) dt^2 + 
\left( 1-\frac{2MG}{r} \right) ^{\! -1} dr^2 +r^2 d \Omega _2 ^2 \, .
\eeq
Euclideanising by $t=- i \tau$ and putting $x=4 M \sqrt{1-2MG/r \,}$, we 
obtain \cite{Haw79}
\beq
	ds^2 = \left( \frac{x}{4MG} \right)^{\! 2} d \tau ^2+ \left( 
	\frac{r^2}{4 M^2 G^2} \right) ^{\! 2} dx^2+r^2 d \Omega _2^2 \, .
\eeq
This metric will be regular at $x=0$, i.e.\ $r=2MG$, if $\tau$ is regarded 
as an angular variable with period $8 \pi MG$. The Euclidean Schwarzschild 
metric is defined on the manifold given by $x \ge 0$, $0 \le 
\tau \le 8 \pi GM$. 
We denote this spacetime as $\calM$ in this section. The submanifold $x=0$ 
is called a bolt \cite{GiH79}, which is the two-dimensional fixed 
point (FP) set 
of the periodic imaginary time isometry. 

Let us consider to calculate the action of $\calM$ by using 
a spacetime $\calM'$ which has 
an outer boundary three-surface $\partial \calM^{\prime}_{\it out}$ at 
infinity and an inner boundary three-surface 
$\partial \calM^{\prime}_{\it in}$ 
at $x=x_{\varepsilon}$ in the vanishing limit of $x_{\varepsilon}$. 
Since the 
metric is a vacuum solution, the Ricci scalar is zero everywhere 
and the action 
$\calS(\calM')$ comes entirely from the boundary terms. 
The Gibbons--Hawking boundary term \cite{Yor, GiH} is given by the integral 
over the boundary
\beq
	\calS_{GH} = -\frac{1}{8 \pi G} \int d^3 \! \xi \sqrt{h} \left( K-K_0 \right) ,
\eeq
where $h$ is the determinant of the induced metric on the boundary, $K$
and $K_0$ are the trace of the extrinsic curvature of the boundary and
the boundary imbedded in flat spacetime, respectively. The second term
is included when we consider a non-compact spacetime. For calculation of the Gibbons--Hawking term it is convenient to use the formula
\beq
	\int d^3 \! \xi \sqrt{h} K = \partial_{normal} (\mbox{Volume of boundary}).
\eeq
The boundary term of the outer boundary $\partial \calM
^{\prime}_{\it out}$
at infinity gives the contribution $4 \pi M^2 G$. 
In addition, the boundary term of the inner boundary $\partial \calM
^{\prime}_{\it in}$ 
is also non-vanishing:
\begin{eqnarray}
    \calS(\partial \calM^{\prime}_{\it in}) &=& \lim_{x_{\varepsilon} \to 0} 
    \calS_{GH}   \\
  	&=& 4 \pi M^2 G .
\end{eqnarray}
But this term is not expected to contribute to the action $\calS(\calM)$ of the
whole spacetime, because the metric is regular at the $x=0$ and so 
the spacetime $\calM$ has no boundary there.
Therefore we expect that the contribution of the excised spacetime 
(in this case, the bolt) will be non-zero and will compensate the 
contribution of the new boundary which is added by considering 
$\calM'$ instead of $\calM$. 

To confirm this, we consider a four-sphere $\calM_B$ which is centered 
around the bolt and whose boundary three-surface is at $x=x_{\varepsilon}$. 
We define $\calS(\mbox{FP})$ as 
the action of $\calM_B$ in the limit of $x_{\varepsilon} \to 0$. Because the 
Euclidean Schwarzschild metric is regular at the bolt, we can evaluate  
this term easily: In the vanishing limit of $x_{\varepsilon}$ the bulk term vanishes and the action comes entirely from the Gibbons--Hawking term of $\partial \calM_B$, and we need not to calculate it because it is given by the Gibbons--Hawking term of $\partial \calM'_{\it in}$ only with the reversal of the direction of the normal of the boundary. So we obtain 
\begin{eqnarray}
	\calS(\mbox{FP}) &=& \lim_{x_{\varepsilon} \to 0} 
    \calS_{GH}   \\
	& = &  -\calS(\partial \calM^{\prime}_{\it in}) .
\end{eqnarray}
The contribution of the bolt cancels with the inner boundary contribution 
of $\calM'$, as expected.
Therefore using $\calM'$ we obtain the correct action of $\calM$ as  
\begin{eqnarray}
	\calS(\calM) &=& \calS(\calM')+\calS(\mbox{FP})  \\
	&=& 4 \pi M^2 G .
\end{eqnarray}

\section{The case of singular instantons} 
$\quad \, $ In this section we calculate the fixed point contribution of 
HT-singularities. The method to calculate it presented in the previous section cannot be applied directly 
because the excised spacetime is not regular in this case. But we show
that we can go
around this obstruction by using Kaluza--Klein technique.

Introduction of an extra dimension in HT-singular instantons was first discussed by Garriga \cite{Gar99, Gar98Apr}.
He showed that Vilenkin's instanton may be 
reinterpreted {\it \`{a} la} Kaluza--Klein as the 
Euclidean Schwarzschild solution of five-dimensional gravity,
and calculated the action of Vilenkin's instanton as the action 
of the Euclidean Schwarzschild solution.
He argued that if the 
fifth dimension is physical and
we can interpret the HT-singularity as such, then the large extra dimension 
as in the M-theory metastabilises the flat spacetime. But at the same time he also argued that the five-dimensional interpretation would not work if a general potential term is present, because HT-singular instantons can be represented as compactified five-dimensional regular metrics only when the initial value of the scalar field takes a particular value. He exemplified this by using a model of five-dimensional gravity with cosmological constant and argued that this 
may indicate that the {\it family} of HT-singular instantons would not exist.

But if we take the HT-singular instantons as `constrained gravitational 
instantons \cite{Wu98},' 
the singularity does not need the justification by five-dimensional regularity. 
Hence in this paper we use the five-dimensional theory as a mere trick only to calculate the contribution of HT-singularity. Then the difficulty mentioned above disappears (see also the discussions in the final section). 
We show that near HT-singularity any HT-singular instantons can be approximated as five-dimensional regular metrics on compactification. It turns out that the HT-singularity corresponds to a bolt in five dimensions. Thus we can calculate the fixed point contribution of HT-singularities by evaluating the bolt contribution in five dimensions. The use of an extra dimension may be seen as a method to constrain HT-singularities.

First we review the behaviour of HT-singular instantons.
The action of the HT-singular instantons is given by 
\beq
  \calS=\int d^4 \! x \sqrt{g} \left( -\frac{\calR}{16 \pi G} +\frac{1}{2}
(\partial \phi)^2 + V(\phi) \right) + \calB ,     \label{4Daction}
\eeq
where $\calB$ represents the boundary contribution $\calS(\mbox{boundary})$ and the fixed point contribution of the HT-singularity $\calS(\mbox{FP})$. 
We seek solutions of $O(4)$ symmetric metric \cite{HaT98Feb}
\beq
  ds^2=d \sigma^2+b^2 ( \sigma) \, d\Omega_3^2 =d \sigma^2+b^2(\sigma) \left( d \psi^2+ \sin ^2 (\psi) \, d\Omega_2^2 \right) .
\eeq
The equations of motion read
\beq
  \phi^{\prime \prime}+3\frac{b^{\prime}}{b} \phi^{\prime} = \frac{\partial V}{\partial \phi} \, , \quad 
  b^{\prime \prime} = -\frac{8 \pi G}{3} b \left( \phi^{\prime \, 2}+V \right)
  \label{Eq_of_Mo}
 \eeq
where primes denote derivatives with respect to $\sigma$.
There are two choices of boundary conditions:

\noindent
{\it 1) Hawking--Turok's case \cite{HaT98Feb}.}
The metric and the scalar field are non-singular at $\sigma=0$, i.e.
\beq
  b(\sigma) \simeq \sigma \quad (\sigma \simeq 0) \, , \quad 
  \phi^{\prime}(0)=0  \, .
\eeq
We can set the value $\phi(0) = \phi_0$ arbitrarily except values which 
correspond to regular instanton solutions. 

\noindent
{\it 2) Asymptotically flat case (including Vilenkin's case \cite{Vil98}).}
\beq
  b(\sigma) \simeq \sigma \, , \quad \phi(\sigma) \to \phi_{\infty} \quad (\sigma
\to \infty) \, ,
\eeq
where $\phi_{\infty}$ represents the value which the field $\phi$ takes at the 
extremum of the potential, $\frac{\partial V}{\partial \phi} (\phi_{\infty})=0$ and
$V(\phi_{\infty})=0$.

Because $b$ goes to zero as we approach the singularity, the (anti-)damping 
term $3b' \phi' /b$ in the equation of motion overwhelms the potential 
term $\partial V/\partial \phi$ 
near singularity, if the potential is not too steep there. 
Hence HT-singular instantons have the similar 
behaviours near the singularity $\sigma=\sigma_*$,
regardless of the potential and the choice of the boundary 
conditions, as
\begin{eqnarray}
	b^3(\sigma) & \simeq & \kappa C \left| \sigma_*-\sigma \right| \label{behave1} \\
	\phi(\sigma) & \simeq & \pm \frac{1}{\kappa} \log \left| \sigma_*-\sigma \right| +{\it const.} \, , \label{behave2}
\end{eqnarray}
where $\kappa=\sqrt{12 \pi G}$ and $C$
is a constant.
We see that the approximation we used is consistent 
if the potential is 
\beq
\mbox{not so steep as } V(\phi) \sim e^{2 \kappa \left| \phi \right| } \quad \mbox{near singularity.}  \label{cond}
\eeq
Using 
Eq.(\ref{behave1}), (\ref{behave2}), (\ref{cond}) and the relation $\calR=6(-b b^{\prime \prime}-b^{\prime \, 2}+1)/b^2$, we can check that the bulk terms of the action Eq.(\ref{4Daction}) do not 
contribute at the singularity.

Secondly let us calculate the boundary contribution $\calS(\mbox{boundary})$ at the singularity.
From the whole spacetime $\calM$ we excise a four-sphere 
which is centered around the singularity and of coordinate radius $\sigma_{\varepsilon}$ and 
call the remaining spacetime as $\calM'$. 
We consider to take the limit of $\sigma_{\varepsilon} \to 0$.
Hereafter we only consider the boundary at singularity. 
The contribution of the boundary at infinity is restored, if any, in the end of the calculation, Eq.(\ref{Conc}). Then
the action contribution of the boundary of $\calM'$
becomes as
\begin{eqnarray}
	\calS(\partial \calM') &=& \lim_{\sigma_{\varepsilon} \to 0} \calS_{GH}  \\
	&=& -\frac{1}{8 \pi G} \left( \epsilon \frac{\partial}{\partial \sigma} 2 \pi^2 b^3(\sigma) \right)_{\sigma \to \sigma_*}  \\ 
	&=& \sqrt{\frac{3 \pi^3}{4G}} \, C \, , \label{GH}
\end{eqnarray}
where $\partial \calM'$ denotes the boundary of $\calM'$ and $\epsilon$ is 
defined by
\beq
	\epsilon=\cases{
		+1 & Hawking--Turok's case  \cr
		-1 & asymptotically flat case.  \cr
		}
\eeq
We see that the constant in the behaviour of $\phi$ (Eq.(\ref{behave2}))
is unimportant for the boundary contribution. This is because the Gibbons--Hawking term involves only $b$, and only $\phi'$ and $b$ are important in the equations of motion near singularity.

Next, we calculate the contribution of the excised spacetime. 
In order to find it, we at first calculate the action contribution of bolts in five dimensions, and then we show the correspondence of the bolts in five-dimensions to HT-singularities in four dimensions.
The five-dimensional action is given by
\beq
  \calS=\int d^5 \! x \sqrt{\tilde{g}} \left( -\frac{1}{16 \pi \tilde{G}} \tilde{\calR} 
     + \mbox{(other terms)} \right)
     -\frac{1}{8 \pi \tilde{G}} \int d^4 \! \xi \sqrt{\tilde{h}} \tilde{K} ,
\eeq
where the tildes distinguish five-dimensional quantities from their 
four-dimen\-sional counterparts. 
We consider $O(4) \times U(1)$ symmetric solutions.
As it turns out below, the submanifold where the fifth dimension closes 
becomes the HT-singularity in four dimensions.
So we approximate the five-dimensional regular metric there as
\beq
    d \tilde{s}^2 
      \simeq d\tau^2+R_0^2 \,
    d \Omega_3^2+\tau^2 \, d \theta^2 \quad (\tau \to 0),
	\label{trial5D}
\eeq
where $R_0$ is a constant and the regularity at $\tau=0$ 
requires that $\theta$ is identified with period $2 \pi$. 
The submanifold $\tau=0$ is the bolt in five dimensions (three-dimensional 
fixed point set).
Its contribution $\calS(\mbox{FP})$ becomes
\begin{eqnarray}
	\calS(\mbox{FP}) &=& \lim_{\tau \to 0} \calS_{GH} \\
	&=& -\frac{1}{8 \pi \tilde{G}} \frac{\partial}{\partial \tau} \left( 2 \pi^2 R^3_0 \cdot 2 \pi \tau \right)_{\tau \to 0}   \\
	&=& -\frac{\pi^2 R_0^3}{2  \tilde{G}}  .
\end{eqnarray}

Then we show the correspondence of the bolts in five dimensions to HT-singularities in four dimensions.
We take fields to be independent of the fifth coordinate $\theta$ and use
the ansatz
\begin{equation}
  \tilde{g}_{AB} = e^{\frac{2 \kappa}{3} ( \pm \phi +a)} \left(
\begin{array}{cc}
g_{\mu \nu} & 0 \\
0 & e^{-2 \kappa (\pm \phi+a)} 
\end{array}
\right)
= \left(
\begin{array}{cc}
  e^{\frac{2 \kappa}{3}(\pm \phi+a)} g_{\mu \nu} & 0 \\
  0 & e^{-\frac{4 \kappa}{3} (\pm \phi+a)}
\end{array}
\right) \, , \label{compact}
\end{equation}
where $a$ is a constant. The addition of $a$ to $\pm \phi$ is a mere shift 
of the zero point 
of $\phi$ and does not affect the calculation of the contribution of singularity to the action, but it is added for clarity.
We obtain the four-dimensional action which has appropriate coefficients
\beq
  \calS=\int d^4 \! x \sqrt{g} \left( -\frac{\calR}{16 \pi G} + \frac{1}{2}
    (\partial \phi)^2 +\mbox {(other terms)} \right)
    -\frac{1}{8 \pi G} \int d^3 \! \xi \sqrt{h} K  .
\eeq
$G$ and $\tilde{G}$ are related by $\tilde{G}=G \int d \theta=2 \pi G$.
From Eq. (\ref{compact}), we see that the four-dimensional fields diverge when the fifth dimension closes. Namely the submanifold where the fifth dimension closes is identified as the singularity in four dimensions. 
The five-dimensional metric of Eq.({\ref{trial5D}}) becomes
\begin{eqnarray}
	\tau^3 & \simeq & \frac{9}{4} \left(\sigma_*-\sigma \right)^2  \\
	b^3(\sigma) & \simeq & \frac{3R_0^3}{2} \left| \sigma_*-\sigma \right|  
	\label{HTfr5D} \\
	\phi(\sigma) & \simeq & \mp \frac{1}{\kappa} \log \left| 
	\sigma_*-\sigma \right| \mp \frac{1}{\kappa} \log 
	\frac{3 e^{\kappa a}}{2} \, ,
\end{eqnarray}
near the bolt, $\tau=0$ i.e.\ $\sigma=\sigma_*$.
We see that any HT-singular behaviours can be obtained by choosing 
$R_0$ and $a$ properly.

Comparing Eq.({\ref{behave1}) and  Eq.(\ref{HTfr5D}), we have 
that $3R_0^3/2=\kappa C$, so the bolt contribution $\calS(\mbox{FP})$ 
can be rewritten in terms of the four-dimensional quantities as 
$\calS(\mbox{FP})=-\sqrt{\pi^3 C^2 /3G}=-\frac{2}{3} \calS(\partial \calM') $.
Hence the total action $\calS(\calM)$ becomes 
\begin{eqnarray}
	\calS(\calM) &=& \calS(\calM')+\calS(\mbox{FP})   \\
	&=& \calS(\mbox{bulk})+\frac{1}{3} \calS(\partial \calM') \, 
	\left( +\calS(\mbox{boundary at infinity}) \right).  \label{Conc}
\end{eqnarray}
Thus the contribution of the HT-singularity to 
the action can be calculated as the one third of the Gibbons--Hawking term
of the three-surface which wraps the singularity. 

\section{The Total Action}    
$\quad \, $ We can obtain a very simple formula for the total 
action of the system considered above. Interestingly, we have the same formula for both the regular and HT-singular instantons only if we use the above results, Eq.(\ref{Conc}). 

From the Einstein equations
\begin{eqnarray}
	b^{\prime \prime} &=& -\frac{8 \pi G}{3} b \left( 
	\phi^{\prime \, 2}+V \right)  \\
	b^{\prime \, 2} &=& \frac{8 \pi G}{3} b^2 \left( \frac{1}{2} 
	\phi^{\prime \, 2}-V \right) +1,
\end{eqnarray}
we have
\beq
	\left( b^3 \right)^{\prime \prime} = 3 \left(-8 \pi G b^3 V+2b \right) .
\eeq
By integration, we obtain the relation
\beq
	\int d\sigma b^3 V=\frac{1}{4 \pi G} \int d\sigma b -\frac{1}{3} \left[ \frac{1}{8 \pi G}(b^3)^{\prime} \right] .
\eeq 
On the other hand, the trace of the Einstein equation reads
\beq
	\calR=8 \pi G \left( (\partial \phi)^2+4V(\phi) \right) .
\eeq
Hence the total action Eq.(\ref{4Daction}) can be rewritten as
\begin{eqnarray}
  \calS &=& -\int d^4 \! x \sqrt{g} V(\phi) +\calB \\
  &=& -2 \pi ^2 \int d \sigma b^3 (\sigma ) V( \phi ) +\calB  \\
	&=& - \frac{\pi}{2G} \int d \sigma b(\sigma) +\frac{1}{3} \left[ \frac{1}{8 \pi G} (2 \pi^2 b^3)^{\prime} \right] +\calB .
\end{eqnarray}

In the case of compact regular instantons, the second and third terms vanish, so we obtain a simple formula
\beq
	\calS=-\frac{\pi}{2G} \int
	d \sigma b.
\eeq
We observe that the total action is given by the area of the circle plotted by $b(\sigma)$ in the $(\sigma, b)$-plane.
In the case of Hawking--Turok instantons, the second term is non-vanishing at the singularity, but it is exactly cancelled by the third term
\beq
	\calB=\frac{1}{3} \left( -\frac{1}{8 \pi G} (2 \pi^2 b^3)^{\prime}\right)_{\sigma \to \sigma_*}. 
\eeq
Therefore we have the same expression for regular and HT-singular 
instantons. Here the coefficient $1/3$ in front of the Gibbons--Hawking 
term which originates from the addition of $\calS(\mbox{FP})$ is crucial. 
It is to be noted that by changing $\phi_0$ we can obtain both regular and 
HT-singular instantons when the potential $V(\phi)$ has additional extrema. 

In the case of HT-singular asymptotically flat instantons, 
the total action becomes 
\begin{eqnarray}
	\calS &=& \lim_{s \to \infty} \left\{ -\frac{\pi}{2G} \int
	_{\sigma_*}^s d\sigma b(\sigma)+\frac{\pi}{12G} ( b^3 )^{\prime} (s) \right\} .
\end{eqnarray}
Here we have dropped the 
boundary contributions at $s$ in the $\calB$ term because they cancels in the limit of $s \to \infty$ on account of the asymptotically flat condition.

\section{Discussions}
$\quad \, $ In conclusion, we showed in this paper that the
HT-singularity has a fixed point contribution to the action and it is
given by $-2/3$ of the Gibbons--Hawking term of the three-surface which
wraps the singularity and whose normal vectors point to the singularity,
in the zero-volume limit of the three-surface. By adding the contribution to the action, we also obtained a simple formula for the action which is feasible for both regular and HT-singular instantons.

Two comments are in order.

At the singularity the equations of motion are not satisfied and so the action is not stationary. But the probability of quantum transition is calculated from the path integral which is integrated over instantons which is constrained by a given 3-metric on the considering three-surface. 
Hence the singularity at the constrained surface would be allowed. 
Motivated from this observation by Wu \cite{Wu98}, it seems plausible that the HT-singularity 
in the cases of the creation of open universe \cite{HaT98Feb} and decay of flat spacetime \cite{Vil98} would be allowed. In these cases the HT-singularity is on the quantum transition surface, so that the parameter $C$ (in other words, $\phi_0$ or $\sigma_*$) is constrained and the dependence of the action on it would be justified. (However, in ref.\ \cite{Wu98} it was suggested to throw away the hemisphere which contains the HT-singularity 
and the whole manifold should be made by joining the remaining regular 
hemisphere and its oriented reversal in order to avoid the trouble caused by the HT-singularity.) Recently, 
Turok \cite{Tur99} introduced a constraint and argued that Vilenkin's instantons may
be defined as a limit of constrained instantons which are non-singular 
(albeit with discontinuous first derivatives) and have no boundary. 
Using such instantons it was argued that the action of Vilenkin's instantons 
is given by the same value calculated by Vilenkin 
\cite{Vil98}, i.e.\ the whole Gibbons--Hawking term. This differs with our result, $1/3$ of the Gibbons--Hawking term, and so it seems doubtful that 
the obtained instantons are Vilenkin's instantons.
The suggested constraint may be an unappropriate one so that 
HT-singularity does not appear from the limit of such instantons. 

As a second comment, 
we examine the five-dimensional theory consisting of gravity and a
cosmological constant $\tilde{\Lambda}$ to show how 
the difficulty 
mentioned in Sec.\ III disappears. In ref.\ \cite{Gar98Apr} 
it was stated that the regular instantons exist in this theory only when
$\phi_0$ is zero. However, careful examination reveals that this is not
the case.
The five-dimensional metric with $O(4) \times U(1)$ symmetry is 
\beq
	d\tilde{s}^2 =d\tau^2+\hat{A}^2 \sin ^2 \left( \frac{\tau}{\hat{A}} \right) d\Omega_3^2+\hat{A}^2 \cos ^2 \left( \frac{\tau}{\hat{A}} \right) d \theta^2 ,
\eeq
where $\hat{A}=\sqrt{6/\tilde{\Lambda}}$. The coordinate $\tau$ runs 
from $0$ to $\hat{A} \pi /2$. This metric is regular everywhere for any $\hat{A}$.
The
HT-singularity appears at $\tau=\hat{A} \pi /2$ on compactification.
The behaviour of the scalar field near $\sigma=0$ (in this case it
corresponds to $\tau=0$) becomes
\beq
  \pm \phi (\sigma) \simeq -\frac{3}{2 \kappa} \log \hat{A}+\frac{3\sigma^2}{4 \kappa
{\hat{A}}^3} - a,   \label{Gar}
\eeq
and the potential becomes
\beq
	V(\phi)=\frac{\tilde{\Lambda}}{8 \pi G} e^{\frac{2 \kappa}{3} 
	(\pm \phi+a)}
	=\frac{3}{4 \pi G} e^{(-2 \log \hat{A}+\frac{2 \kappa}{3} a)} \, 
	e^{\pm \frac{2 \kappa}{3} \phi} \, .
\eeq
From these equations we see that we can have any $\phi_0
$ without changing the four-dimensional potential term by choosing $a$ and $\tilde{\Lambda}$ appropriately. 

We may interpret this calculation 
in terms of constraints on instantons as follows. Four-dimensional singular instantons
are one parameter family in constrast to regular instantons and the
action depends on the parameter. This parameter enters the instanton solution
as an integration constant of the equations of motion. 
We saw in Sec.\ III that regular instantons in five dimensions whose fifth
dimension closes show the HT-singular behaviour on compactification. 
In general there would be only a finite number of such regular instantons in a
five-dimensional theory.  
They have one-to-one correspondence to four-dimensional 
HT-singular instantons of different parameter, so that we have a 
constraint on the parameter of the HT-singular instantons. 
If we choose one five-dimensional theory (i.e.\ the value of $\tilde{\Lambda}$, in the above example), then one or some of HT-singular instantons are picked out from the family of HT-singular instantons. 
The parameter of the HT-singular instantons
can be later integrated over in path integral by integrating over the appropriate parameter space of the constraining five-dimensional theory. 

\section*{Acknowledgements}
 I am grateful to Ken-Iti Izawa and Hiroaki Terashima for reading the 
 manu\-script and for useful comments.


\newpage

\begin{thebibliography}{99}

\bibitem{HaT98Feb}
S. W. Hawking and N. Turok, Phys. Lett. {\bf B425}, 25 (1998).

\bibitem{TuH98}
N. Turok and S. W. Hawking, Phys. Lett. {\bf B432}, 271 (1998).

\bibitem{HaH83}
J. B. Hartle and S. W. Hawking, Phys. Rev. D {\bf 28}, 2960 (1983).

\bibitem{Gar98Mar}
J. Garriga, Phys. Rev. D {\bf 61}, 047301 (2000).

\bibitem{GMST98}
J. Garriga, X. Montes, M. Sasaki and T. Tanaka, Nucl. Phys. 
{\bf B513}, 343 (1998).

\bibitem{Gar99}
J. Garriga, Int. J. Theor. Phys. {\bf 38}, 2959 (1999).

\bibitem{Wu98}
Z. C. Wu, Gen. Rel. Grav. {\bf 30}, 1639 (1998).

\bibitem{Vil98}
A. Vilenkin, Phys. Rev. D {\bf 57}, 7069 (1998).

\bibitem{Gar98Apr}
J. Garriga, ``Smooth Creation of an Open Universe in Five Dimensions,'' preprint UAB-FT-441, hep-th/9804106.

\bibitem{Wit82}
E. Witten, Nucl. Phys. {\bf B195}, 481 (1982).

\bibitem{Tur99}
N. Turok, Phys. Lett. {\bf B458}, 202 (1999).

\bibitem{BoL98}
R. Bousso and A. Linde, Phys. Rev. D {\bf 58}, 083503 (1998).

\bibitem{Yor}
J. W. York, Jr., Phys. Rev. Lett. {\bf 28}, 1082 (1972).

\bibitem{GiH}
G. W. Gibbons and S. W. Hawking, Phys. Rev. D {\bf 15}, 2752 (1977).

\bibitem{BoC}
R. Bousso and A. Chamblin, Phys. Rev. D {\bf 59}, 063504 (1999).

\bibitem{Gon98}
P. F. Gonz\'{a}lez-D\'{\i}az, Phys. Rev. D {\bf 59}, 043509 (1999).

\bibitem{Haw79}
S. W. Hawking, in {\it General Relativity: An Einstein Centenary Survey}, 
edited by S. W. Hawking and W. Israel (Cambridge University Press, Cambridge, England, 1979).

\bibitem{GiH79}
G. W. Gibbons and S. W. Hawking, Commun. Math. Phys. {\bf 66}, 291 (1979).

\end{thebibliography}
\end{document}